\RequirePackage[pagewise,mathlines]{lineno}

\documentclass[twocolumn,aps,showpacs,superscriptaddress]{revtex4-1}
\usepackage{savesym}
\usepackage{amssymb}
\usepackage{xcolor}
\usepackage{bbding}
\usepackage{multirow}
\usepackage{enumerate}
\usepackage{graphicx}

\usepackage{dcolumn}
\restoresymbol{TXT}{affil}
\usepackage{bm}

\begin{document}

\def\Journal#1#2#3#4{{#1} {\bf #2}, #3 (#4)}

\def\NCA{Nuovo Cimento}
\def\NIM{Nucl. Instr. Meth.}
\def\NIMA{{Nucl. Instr. Meth.} A}
\def\NPB{{Nucl. Phys.} B}
\def\NPA{{Nucl. Phys.} A}
\def\PLB{{Phys. Lett.}  B}
\def\PRL{Phys. Rev. Lett.}
\def\PRC{{Phys. Rev.} C}
\def\PRD{{Phys. Rev.} D}
\def\ZPC{{Z. Phys.} C}
\def\JPG{{J. Phys.} G}
\def\CPC{Comput. Phys. Commun.}
\def\EPJ{{Eur. Phys. J.} C}
\def\PR{Phys. Rept.}
\def\PRV{Phys. Rev.}
\def\JHEP{JHEP}

\preprint{}
\title{Energy dependence of $J/\psi$ production in Au+Au collisions at $\sqrt{s_{NN}} =$ 39, 62.4 and 200 GeV }
\date{\today}
\affiliation{AGH University of Science and Technology, FPACS, Cracow 30-059, Poland}
\affiliation{Argonne National Laboratory, Argonne, Illinois 60439}
\affiliation{Brookhaven National Laboratory, Upton, New York 11973}
\affiliation{University of California, Berkeley, California 94720}
\affiliation{University of California, Davis, California 95616}
\affiliation{University of California, Los Angeles, California 90095}
\affiliation{Central China Normal University, Wuhan, Hubei 430079}
\affiliation{University of Illinois at Chicago, Chicago, Illinois 60607}
\affiliation{Creighton University, Omaha, Nebraska 68178}
\affiliation{Czech Technical University in Prague, FNSPE, Prague, 115 19, Czech Republic}
\affiliation{Nuclear Physics Institute AS CR, 250 68 Prague, Czech Republic}
\affiliation{Frankfurt Institute for Advanced Studies FIAS, Frankfurt 60438, Germany}
\affiliation{Institute of Physics, Bhubaneswar 751005, India}
\affiliation{Indian Institute of Technology, Mumbai 400076, India}
\affiliation{Indiana University, Bloomington, Indiana 47408}
\affiliation{Alikhanov Institute for Theoretical and Experimental Physics, Moscow 117218, Russia}
\affiliation{University of Jammu, Jammu 180001, India}
\affiliation{Joint Institute for Nuclear Research, Dubna, 141 980, Russia}
\affiliation{Kent State University, Kent, Ohio 44242}
\affiliation{University of Kentucky, Lexington, Kentucky, 40506-0055}
\affiliation{Korea Institute of Science and Technology Information, Daejeon 305-701, Korea}
\affiliation{Institute of Modern Physics, Chinese Academy of Sciences, Lanzhou, Gansu 730000}
\affiliation{Lawrence Berkeley National Laboratory, Berkeley, California 94720}
\affiliation{Lehigh University, Bethlehem, PA, 18015}
\affiliation{Max-Planck-Institut fur Physik, Munich 80805, Germany}
\affiliation{Michigan State University, East Lansing, Michigan 48824}
\affiliation{National Research Nuclear Univeristy MEPhI, Moscow 115409, Russia}
\affiliation{National Institute of Science Education and Research, Bhubaneswar 751005, India}
\affiliation{National Cheng Kung University, Tainan 70101 }
\affiliation{Ohio State University, Columbus, Ohio 43210}
\affiliation{Institute of Nuclear Physics PAN, Cracow 31-342, Poland}
\affiliation{Panjab University, Chandigarh 160014, India}
\affiliation{Pennsylvania State University, University Park, Pennsylvania 16802}
\affiliation{Institute of High Energy Physics, Protvino 142281, Russia}
\affiliation{Purdue University, West Lafayette, Indiana 47907}
\affiliation{Pusan National University, Pusan 46241, Korea}
\affiliation{Rice University, Houston, Texas 77251}
\affiliation{University of Science and Technology of China, Hefei, Anhui 230026}
\affiliation{Shandong University, Jinan, Shandong 250100}
\affiliation{Shanghai Institute of Applied Physics, Chinese Academy of Sciences, Shanghai 201800}
\affiliation{State University Of New York, Stony Brook, NY 11794}
\affiliation{Temple University, Philadelphia, Pennsylvania 19122}
\affiliation{Texas A\&M University, College Station, Texas 77843}
\affiliation{University of Texas, Austin, Texas 78712}
\affiliation{University of Houston, Houston, Texas 77204}
\affiliation{Tsinghua University, Beijing 100084}
\affiliation{United States Naval Academy, Annapolis, Maryland, 21402}
\affiliation{Valparaiso University, Valparaiso, Indiana 46383}
\affiliation{Variable Energy Cyclotron Centre, Kolkata 700064, India}
\affiliation{Warsaw University of Technology, Warsaw 00-661, Poland}
\affiliation{Wayne State University, Detroit, Michigan 48201}
\affiliation{World Laboratory for Cosmology and Particle Physics (WLCAPP), Cairo 11571, Egypt}
\affiliation{Yale University, New Haven, Connecticut 06520}

\author{L.~Adamczyk}\affiliation{AGH University of Science and Technology, FPACS, Cracow 30-059, Poland}
\author{J.~K.~Adkins}\affiliation{University of Kentucky, Lexington, Kentucky, 40506-0055}
\author{G.~Agakishiev}\affiliation{Joint Institute for Nuclear Research, Dubna, 141 980, Russia}
\author{M.~M.~Aggarwal}\affiliation{Panjab University, Chandigarh 160014, India}
\author{Z.~Ahammed}\affiliation{Variable Energy Cyclotron Centre, Kolkata 700064, India}
\author{I.~Alekseev}\affiliation{Alikhanov Institute for Theoretical and Experimental Physics, Moscow 117218, Russia}
\author{A.~Aparin}\affiliation{Joint Institute for Nuclear Research, Dubna, 141 980, Russia}
\author{D.~Arkhipkin}\affiliation{Brookhaven National Laboratory, Upton, New York 11973}
\author{E.~C.~Aschenauer}\affiliation{Brookhaven National Laboratory, Upton, New York 11973}
\author{M.~U.~Ashraf}\affiliation{Tsinghua University, Beijing 100084}
% ...

\collaboration{STAR Collaboration}\noaffiliation
\begin{abstract}
The inclusive $J/\psi$ transverse momentum ($p_{T}$) spectra and nuclear modification factors are reported at midrapidity ($|y|<1.0$) in Au+Au collisions at $\sqrt{s_{NN}}=$ 39, 62.4 and 200 GeV taken by the STAR experiment. A suppression of $J/\psi$ production, with respect to {\color{black}the production in $p+p$ scaled by the number of binary nucleon-nucleon collisions}, is observed in central Au+Au collisions at these three energies. No significant energy dependence of nuclear modification factors is found within uncertainties. The measured nuclear modification factors can be described by model calculations that take into account both suppression of direct $J/\psi$ production due to the color screening effect and $J/\psi$ regeneration from  recombination of uncorrelated charm-anticharm quark pairs.

\end{abstract}
\maketitle

\section{Introduction}
The Relativistic Heavy Ion Collider (RHIC) was built to investigate strongly interacting matter at high temperature and energy density in the laboratory through high-energy heavy-ion collisions. At extremely high temperatures and baryon densities, a transition from the hadronic phase of matter to a new deconfined partonic phase, the Quark-Gluon Plasma (QGP), is predicted by Quantum Chromodynamics (QCD) ~\cite{Harrison_Nucl}. {\color{black}It has been proposed that the color potential in quarkonia  could be screened by quarks and gluons in the QGP ~\cite{Matsui_Phys}. Quarkonia are bound states of charm-anticharm ($c\bar{c}$) or bottom-antibottom ($b\bar{b}$) quark pairs.} As a consequence, quarkonium production cross sections in heavy-ion collisions divided by the corresponding number of binary nucleon-nucleon collisions, $N_{coll}$, are expected to be suppressed compared to those in $p+p$ collisions if QGP is formed in heavy-ion collisions.

The $J/\psi$ is the most abundantly produced quarkonium state accessible to experiments. Over the past twenty years, $J/\psi$ suppression in hot and dense media has been a topic of growing interest. Various measurements of $J/\psi$ have been performed in different collision systems and at different energies, and indeed a suppression of $J/\psi$ production has been observed ~\cite{Lansberg_Mod,Abreu_phys,Arnaldi_Phys,Atomssa_Eur}. A similar centrality dependent suppression was found at SPS (S+U $\sqrt{s_{NN}}$ = 19.4 GeV ~\cite{SPS_jpsi1}, Pb+Pb $\sqrt{s_{NN}}$ = 17.2 GeV ~\cite{SPS_jpsi2} and In+In $\sqrt{s_{NN}}$ = 17.2 GeV ~\cite{Arnaldi_Phys}) and at RHIC (Au+Au $\sqrt{s_{NN}}$ = 200 GeV ~\cite{Adare_Phys,jpsi_chris}) for midrapidity, even though the temperature and energy density reached in these studies are significantly different ~\cite{energy_density}. Furthermore, a stronger suppression at forward rapidity {\color{black}($1.2<|y|<2.2$)} compared to midrapidity {\color{black}($|y|<0.35$)} was observed at RHIC ~\cite{Adare_Phys}. These observations indicate that effects other than color screening are important for $J/\psi$ production. Among these effects, $J/\psi$ production from the recombination of $c\bar{c}$  ~\cite{10_zebo} was suggested to explain the suppressions at SPS and RHIC ~\cite{Zhao_Phys}. With the higher temperature and density at RHIC, the increased contribution due to regeneration from the larger charm quark density could compensate for the enhanced suppression. This could also explain a stronger suppression at forward rapidity at RHIC where the charm quark density is lower compared to midrapidity ~\cite{Zhao_Phys, 10_zebo, Hot_2, Hot_3}. In addition to the color screening and regeneration effects, there are also modifications from cold nuclear matter (CNM) effects and other final state effects, such as nuclear parton distribution function modification ~\cite{CNM_1}, initial energy loss ~\cite{CNM_2}, Cronin effect ~\cite{CNM_3}, nuclear absorption ~\cite{CNM_4} and dissociation by co-movers ~\cite{CNM_5}. The suppression due to these effects has been systematically studied experimentally via $p+$A collisions ~\cite{CNME_1,CNME_2,CNME_3,CNME_4,CNME_5,CNME_6,CNME_7,CNME_8,CNME_9}.  However, the extrapolation from $p+$A to A+A is still model dependent. 

The nuclear modification factor of $J/\psi$ production in Pb-Pb collisions at $\sqrt{s_{NN}} = 2.76$ TeV has been measured at the LHC ~\cite{Jpsi_ALICE,Jpsi_CMS,Jpsi_ATLAS}. In comparison with results from RHIC in Au$+$Au collisions at $\sqrt{s_{NN}} = $ 200 GeV, the $J/\psi$ production is significantly less suppressed, which suggests significantly more recombination contribution at LHC energies. The measurement of $J/\psi$ production at forward rapidity ($1.2<|y|<2.2$) in Au+Au collisions by the PHENIX experiment at $\sqrt{s_{NN}} = $ 39 and 62.4 GeV indicates a similar suppression level as that at $\sqrt{s_{NN}} =$ 200 GeV ~\cite{Jpsi_phenix}. Measurements of $J/\psi$ invariant yields at different collision energies at RHIC in different centralities at mid-rapidity can shed new light on the interplay of these mechanisms for $J/\psi$ production and properties of the medium.

In this letter, we further study the collision energy dependence of $J/\psi$ production and test the hypothesis of these two competing mechanisms of color screening and regeneration in the hot medium. We present  measurements of the $J/\psi$ production at midrapidity ($|y| < 1$) with the STAR experiment in Au+Au collisions at $\sqrt{s_{NN}}$= 39, 62.4 and 200 GeV using data collected during 2010 and 2011 running at RHIC and study the nuclear modification factors at these energies. The data sample used in this analysis (RHIC Run 2011) is different from the previous published results ~\cite{jpsi_chris} (RHIC Run 2010) for Au$+$Au collisions at  $\sqrt{s_{NN}}$= 200 GeV.

\section{Experiment and Analysis}
The STAR experiment is a large-acceptance multi-purpose detector which covers full azimuth with pseudorapidity of $|\eta|<1$ ~\cite{Ackermannn_Nucl}.  The Vertex Position Detector (VPD) was used to select Au+Au collisions that were within {\color{black}$\pm$}15 cm of the center of the STAR detector ~\cite{VPD}. The total numbers of 0-60$\%$ central minimum-bias events that are used in this analysis are 182 million, 94 million, and 360 million for 39, 62.4 and 200 GeV, respectively. The $J/\psi$ is reconstructed through its decay into electron-positron pairs, $J/\psi \rightarrow e^{+}e^{-}$ (branching ratio {\color{black}Br($J/\psi \rightarrow e^{+}e^{-}$)}= 5.97$\pm$ 0.03$\%$ ~\cite{PDG}). The primary detectors used in this analysis are the Time Projection Chamber (TPC) ~\cite{Anderson_Nucl}, the Time-of-Flight (TOF) detector ~\cite{Llope_Nucl}, and the Barrel Electromagnetic Calorimeter (BEMC) ~\cite{Beddo_Nucl}. The TPC provides tracking and particle identification via the ionization energy loss ($\langle dE/dx\rangle$) of charge particles. The TOF ~\cite{Llope_Nucl} measures the velocity of particles, which greatly improved electron identification at low $p_{T}$. This detector, combined with the TPC ~\cite{Anderson_Nucl},  clearly identifies electrons  by rejecting  hadrons in the  low and intermediate $p_{T}$ range {\color{black}($p_{T} < 1.5$ GeV/$c$)}. The BEMC ~\cite{Beddo_Nucl}, a lead-scintillator calorimeter, {\color{black}is} used to improve the electron identification at high $p_{T}$ {\color{black}($p_{T} > 1.5$ GeV/$c$)}. The electron identification method is similar to Ref.~\cite{jpsi_chris, jpsi_zebo}.

Collision centrality was determined from the uncorrected charged particle multiplicity $dN/d\eta$ within $|\eta| < 0.5$ using a Monte Carlo (MC) Glauber model ~\cite{Glauber_model}. The dependence of $dN/d\eta$ on the collision vertex position $V_{z}$ and the beam luminosity has been included to take acceptance and efficiency changes on the measured $dN/d\eta$ into account. For each collision centrality, an average impact parameter, $\langle b \rangle$, average number of participants, $\langle N_{part} \rangle$, and average number of binary collisions, $\langle N_{coll} \rangle$, were related to an observed multiplicity range. Centrality definitions in Au$+$Au collisions for $\sqrt{s_{NN}} = $ 39, 62.4 and 200 GeV are summarized in Table ~\ref{table0}.

 \renewcommand{\floatpagefraction}{0.75}
\begin{table}[htbp]
\newcommand{\tabincell}
\centering
\begin{center}
\begin{tabular}{ccccc}
\hline
\hline
$\sqrt{s_{NN}}$ (\text{GeV})&Centrality ($\%$) & $\langle N_{part}\rangle$ & $\langle N_{coll}\rangle$ & $\langle b \rangle (\text{fm})$ \\
\hline
\multirow{4}{*}{39}&
0 - 20  & 273 $\pm$ 6&629 $\pm$ 26  & 4.4 $\pm$ 0.2\\&
20 - 40 & 137 $\pm$ 11&245 $\pm$ 26 & 8.0 $\pm$ 0.3\\&
40 - 60 & 59 $\pm$ 10&79 $\pm$ 17&  10.4 $\pm$0.4\\&
0 - 60 &156 $\pm$ 8&316 $\pm$ 22 & 7.6 $\pm$ 0.3\\
\hline
\hline
\multirow{4}{*}{62}&
0 - 20  & 276 $\pm$ 5&664 $\pm$ 25  & 4.4 $\pm$ 0.2\\&
20 - 40 & 139 $\pm$ 10&258 $\pm$ 27 & 8.0 $\pm$ 0.3\\&
40 - 60 & 60 $\pm$ 10&82 $\pm$ 18&  10.4 $\pm$0.4\\&
0 - 60 &157 $\pm$ 9&332 $\pm$ 23 & 7.6 $\pm$ 0.3\\
\hline
\hline
\multirow{4}{*}{200}&
0 - 20  & 280 $\pm$ 6&785 $\pm$ 29  & 4.4 $\pm$ 0.2\\&
20 - 40 & 142 $\pm$ 11&300 $\pm$ 31 & 8.0 $\pm$ 0.3\\&
40 - 60 & 62 $\pm$ 10&95 $\pm$ 21&  10.4 $\pm$0.4\\&
0 - 60 &161 $\pm$ 9&393 $\pm$ 27 & 7.6 $\pm$ 0.3\\
\hline
\hline
\end{tabular}
\end{center}
\caption{Summary of centrality bins, average number of participants $\langle N_{part} \rangle$, number of binary collisions $\langle N_{coll} \rangle$, and impact parameter $\langle b \rangle$ from MC Glauber simulation of Au$+$Au at $\sqrt{s_{NN}} = 39$, 62 and 200 GeV. The errors indicate uncertainties from the MC Glauber calculations.}
\label{table0}
\end{table}

The daughter tracks of the $J/\psi$ {\color{black}candidates} are required to have at least 25 out of the 45 possible TPC hits, and a distance of closest approach ({\color{black}DCA}) from the primary vertex {\color{black}of} less than 3 cm. Low momentum ($p < 1.5$ GeV/$c$) electron and positron candidates are  separated from hadrons by selecting on the inverse velocity, {\color{black}$|{1/\beta}-1|<0.03$}, where $\beta$ is the velocity measured in the TOF normalized by the speed of light.  The cut value is determined using a three standard deviation window. At high momentum ($p > 1.5$ GeV/$c$), a cut on the ratio of momentum to energy deposited in towers from BEMC  ($0.3<pc/E<1.5$) is used to suppress hadrons. The electron and positron candidates are then identified by their specific energy loss ($\langle dE/dx\rangle$) in the TPC. More than 15 TPC hits are required to calculate $\langle dE/dx\rangle$. The normalized $\langle dE/dx\rangle$ is defined as follows:\begin{equation}
\label{eq_nse}
n\sigma_{e} = \frac{ln(\langle dE/dx \rangle^{m}/\langle dE/dx \rangle^{th}_{e})}{R_{dE/dx}}
\end{equation}
where $\langle dE/dx \rangle^{m}$ and $\langle dE/dx \rangle^{th}$ represent measured and theoretical values, respectively, and $R_{dE/dx}$ is the experimental $ln(dE/dx)$ resolution. The $n\sigma_{e}$ cut for electron identification is  $ -1.5< n\sigma_{e} < 2$. The combination of these cuts enables the identification of electrons and positrons over a wide momentum range ~\cite{jpsi_chris}. The electron sample purity integrated over the measured $p_{T}$ region is over 90$\%$. Our measurement of $J/\psi$ covers the rapidity range $|y|<1$ due to the STAR acceptance and decay kinematics.

The $J/\psi$ signal is extracted by subtracting combinatorial background reconstructed from {\color{black}the} unlike-sign mixed-events spectrum. The like-sign and mixed-events distributions are obtained as follows:
\begin{enumerate}[1)] \setlength{\itemsep}{0pt}
\item Like-sign:  Electrons (or positrons) of the same charge sign are paired within the same event.
\item {\color{black}Mixed-events}: Events are categorized according to the position along the beam line of the primary vertex  and centrality of the event. Electrons from one event are paired with positrons from other random events from an event pool with similar global features such as collision centrality and vertex position. The vertex position is divided  into 20 bins and the event centrality into 10 bins to ensure that the mixing is done using tracks from similar {\color{black}conditions}.
\end{enumerate}
\renewcommand{\floatpagefraction}{0.75}
\begin{figure*}[htbp]
\includegraphics[angle=0,keepaspectratio,width=0.7\textwidth]{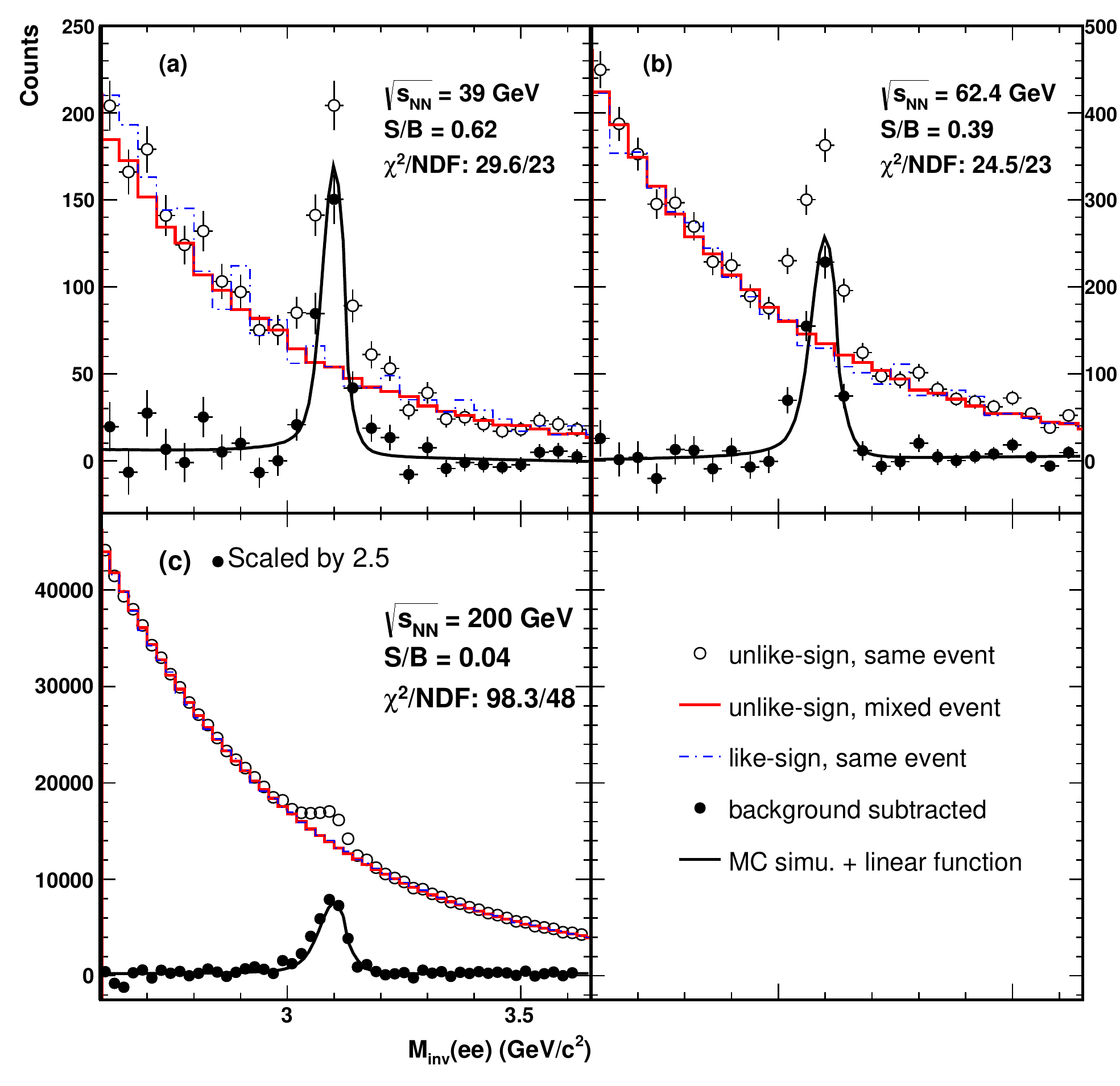}
\caption{ The $e^{+}e^{-}$ invariant mass distribution of $J/\psi$ candidates (black open {\color{black}circles}), like-sign combinatorial background (blue dashed line), mixed event combinatorial background (red solid line), and $J/\psi$ candidates with mixed event background subtracted (black solid  {\color{black}circles}) in Au+Au collisions at $\sqrt{s_{NN}}$ = 39 (a), 62.4 (b), and 200 GeV (c)  for centrality 0 - 60 $\%$.  The $J/\psi$ signal shape from a MC simulation is combined with a linear residual background and is fitted to the combinatorial background subtracted data (black solid line).}
\label{jpsi_signal_Fig}
\end{figure*}

The invariant mass distribution of $e^{+} e^{-}$ pairs before and after the combinatorial background subtraction in 0 - 60 $\%$ central Au+Au collisions are shown in Fig.~\ref{jpsi_signal_Fig} for $\sqrt{s_{NN}}$ = 39, 62.4, and 200 GeV. The mixed-event background is normalized to the like-sign distribution in a mass range of 2.0 - 4.0 GeV/$c^{2}$ and the normalized shapes show close agreement. For the results reported in this paper, we use the mixed-event method for the combinatorial background subtraction. The mass distribution of  $e^{+}e^{-}$ is fitted by $J/\psi$ signal shape obtained from MC simulation, which includes the resolution of the TPC and bremsstrahlung of the daughter electrons in the detector, combined with a straight line for residual background. The residual background mainly comes from the correlated open charm decays and Drell-Yan processes. The raw $J/\psi$ signal is obtained from bin counting in the mass range 2.7 - 3.2 GeV/$c^{2}$ after combinatorial and residual background subtraction. The fraction of $J/\psi$ counts outside of the mass window was determined from the $J/\psi$ MC simulated signal shape and was found to be $\sim$ 5$\%$. This was used to correct the number of $J/\psi$ counts. The modified $J/\psi$ signal shape due to internal radiation was also considered and has been treated as a source of systematic uncertainties ($\sim$ 5$\%$) in yield extraction. Signal-to-background ratios for these three energies are observed to be 0.62, 0.39, and 0.04, respectively for $0\!<\!p_T\!<\!3$ GeV/$c$ (39 and 62 GeV) and $0\!<\!p_T\!<\!5$ {\color{black}GeV/$c$} (200 GeV). 
The $J/\psi$ invariant yield is defined as
\begin{equation}
\label{eq_invariant}
{\color{black}\text{Br}_{J/\psi \rightarrow e^{+}e^{-}}}\frac{d^{2}N}{2 \pi p_{T} dp_{T}dy}
 = \frac{1}{2{\pi}p_{T}{\Delta}p_{T}{\Delta}y} \frac{N_{J/{\psi} \rightarrow e^{+}e^{-}}}{A{\epsilon} N_{EVT}}
\end{equation}
where $N_{J/\psi}$ is the uncorrected number of reconstructed $J/\psi$, $N_{EVT}$ is the number of events in the relevant Au$+$Au centrality selection, A$\epsilon$ is the detector's geometric acceptance times its efficiency (about 0.05 $\sim$ 0.12 for 0 $< p_{T} <$ 5 GeV/$c$), and ${\Delta}p_{T}$
 and ${\Delta}y$ are the bin width in $p_{T}$ and $y$, respectively. Acceptance and efficiency corrections (TPC and BEMC related) are estimated by {\color{black}MC simulations with GEANT3 package} ~\cite{19_zebo}. Some of the efficiency corrections such as TOF and $dE/dx$ related cuts are extracted directly from data ~\cite{dielectron}.
 
The systematic uncertainty on the efficiency correction obtained from MC simulations is estimated by comparing the difference for the particle identification cut distributions between simulation and data. In order to account for the contributions from radiation losses and correlated background in yield extraction procedure, the mass window and methods for signal counting have also been varied to evaluate the uncertainties. The total systematic uncertainties in the integrated $p_{T}$ range are $20\%$, $11\%$, and $10\%$ at $\sqrt{s_{NN}} =$ 39, 62.4, and 200 GeV, respectively. Table~\ref{table1} contains a summary of  the  contributions from the different sources. The agreement of distributions, related to BEMC cuts, between data and MC simulations at $\sqrt{s_{NN}}$ = 39 GeV is less precise owing to the large uncertainty of the BEMC related cuts. The centrality and transverse momentum dependence of the total systematic uncertainties are reflected in the results shown in Section \uppercase\expandafter{\romannumeral3}.

 \renewcommand{\floatpagefraction}{0.75}
\begin{table}[htbp]
\newcommand{\tabincell}
\centering
\begin{center}
\begin{tabular}{cccc}
\hline
Systematic uncertainty source & 39 GeV & 62.4 GeV & 200 GeV \\
\hline
TPC tracking cuts ($\%$) & 8 & 7 & 6\\
BEMC related cuts ($\%$) & 17-25 & 3-5 & 1-2\\
TOF related cuts ($\%$) & 2  & 2 & 2\\
Yield extraction ($\%$) &6-12 & 2-7& 5-11\\
Total ($\%$) & 19-29 &10-12&8-12\\
\hline
\end{tabular}
\end{center}
\caption{The contributions of systematic uncertainty sources for 39{\color{black}, 62.4 and 200 GeV}.}
\label{table1}
\end{table}

\section{Results}
\begin{figure*}[htbp]
\includegraphics[angle=0,keepaspectratio,width=1.0\textwidth]{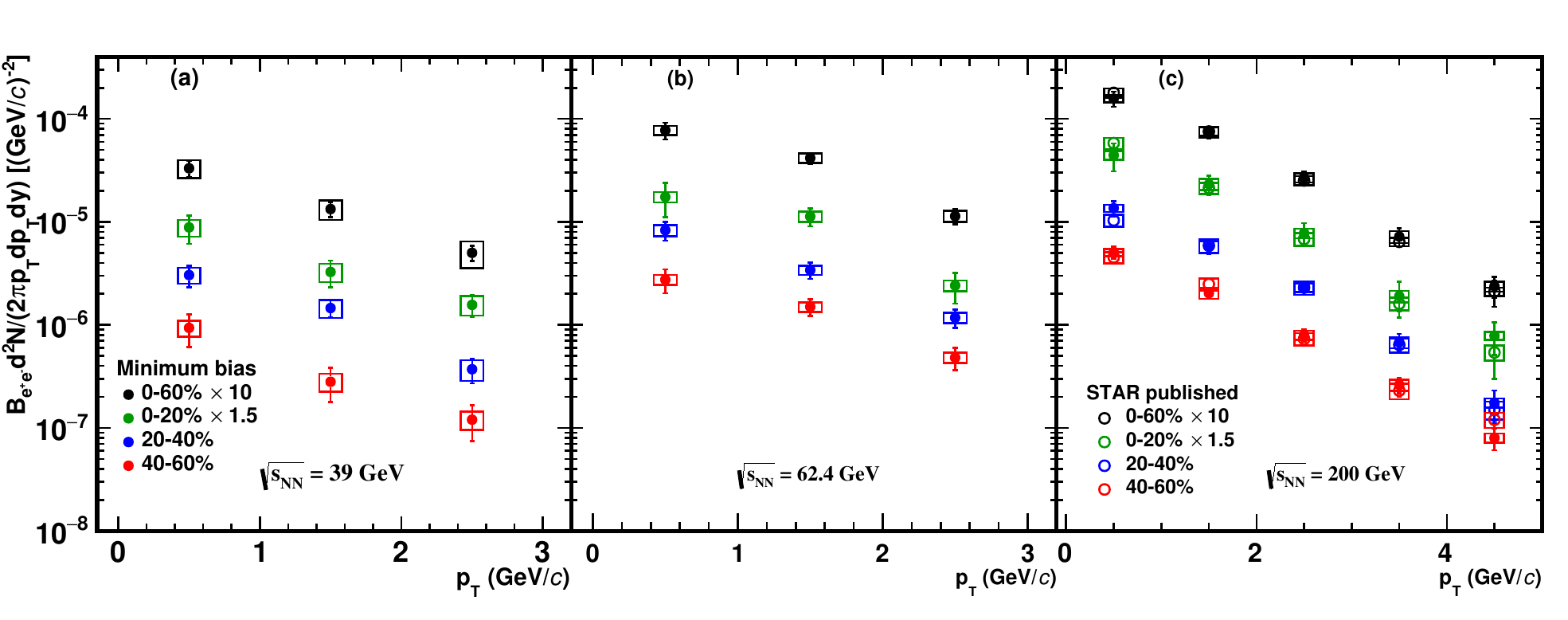}
\caption{ $J/\psi$ invariant yields in Au+Au collisions at $\sqrt{s_{NN}}$ = 39, 62.4 and 200 GeV as a function of $p_{T}$ for different centralities. The error bars represent the  statistical uncertainties. The boxes represent the systematic uncertainties. The STAR published results are from Refs.~\cite{jpsi_zebo} and ~\cite{jpsi_chris}.}
\label{jpsi_spectrum_Fig}
\end{figure*}

The $J/\psi$ invariant yields as a function of $p_{T}$ in Au+Au collisions at $\sqrt{s_{NN}} = $ 39, 62.4, and 200 GeV for different centrality bins are shown in Fig.~\ref{jpsi_spectrum_Fig}. As expected, the $J/\psi$ invariant yields are larger in Au+Au collisions at larger center-of-mass energies. Results from the current measurements (year 2011) are compared with the published results from data taken in 2010.

\renewcommand{\floatpagefraction}{0.75}

\renewcommand{\floatpagefraction}{0.75}
\begin{figure*}[htbp]
\includegraphics[angle=0,keepaspectratio,width=0.45\textwidth]{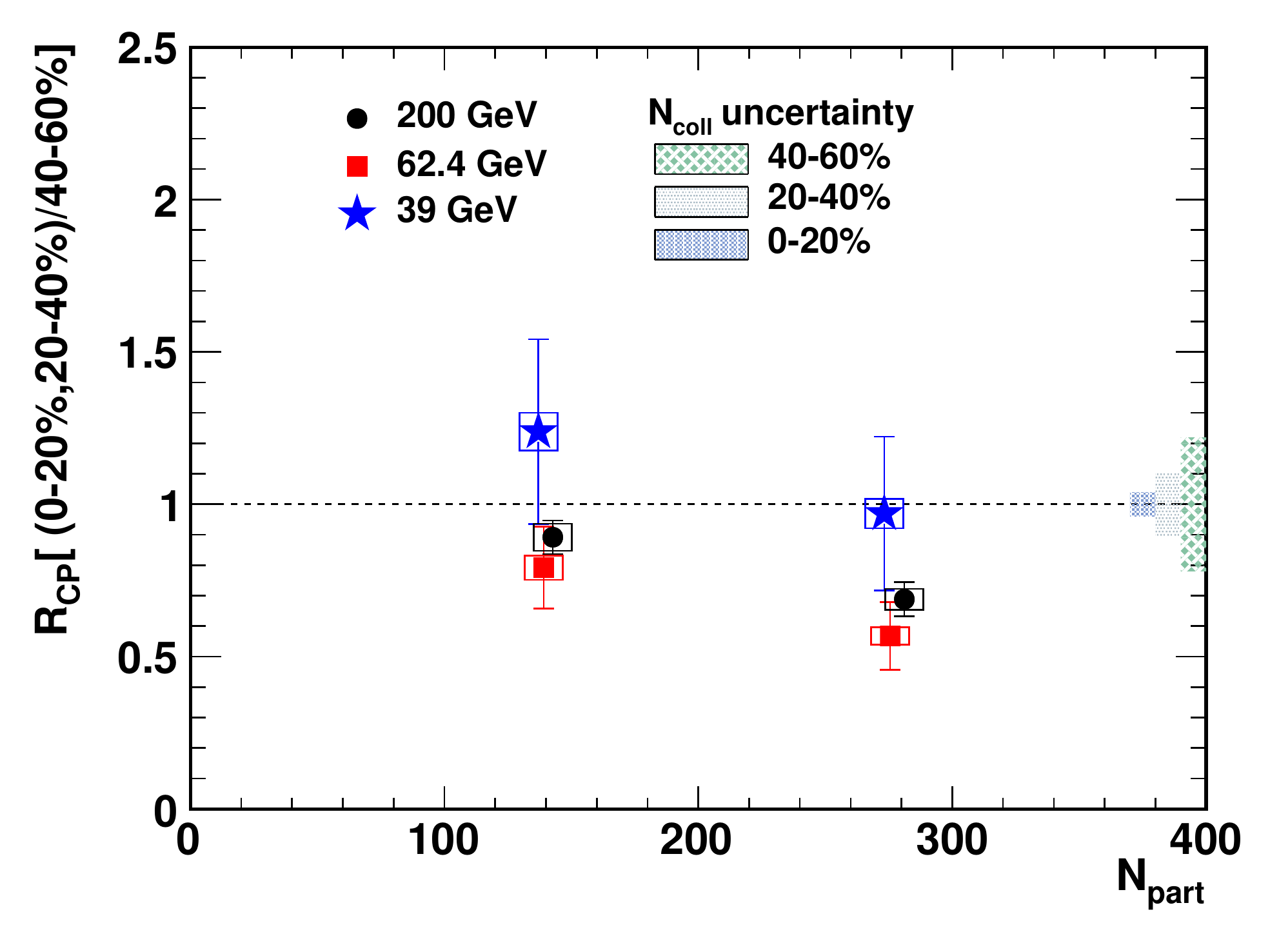}
\caption{$J/\psi$ $R_{CP}$ results (with respect to $40 - 60\%$ peripheral) for Au+Au collisions  as a function of $N_{part}$. The error bars represent the statistical uncertainties.  The boxes represent the systematic uncertainties. The shaded bands represent the normalization uncertainty from $\langle N_{coll}\rangle$ in different centrality bins.}
\label{rcp_Fig}
\end{figure*}

Nuclear modification factors ($R_{CP}$, $R_{AA}$) are used to quantify the suppression of $J/\psi$ production. $R_{CP}$ is a ratio of the $J/\psi$ yield in central collisions to peripheral collisions (centrality: 40-60$\%$) and  defined as follows:\\
\begin{equation}
\label{eq1}
R_{CP}=\frac{\frac{dN/dy}{\langle N_{coll}\rangle}(central)}{\frac{dN/dy}{\langle N_{coll}\rangle}(peripheral)}
\end{equation}
\\where $\langle N_{coll}\rangle$ and $\frac{dN/dy}{\langle N_{coll}\rangle}$ are the average number of nucleon-nucleon collisions and  $J/\psi$ yield per nucleon-nucleon collision in a given centrality, respectively.  $dN/dy$ is obtained {\color{black}from} the integration of the $J/\psi$ $p_{T}$ spectrum. Due to the limited $p_{T}$ coverage of the measurements, the extrapolation of the $p_{T}$ spectrum is done by the two following functions:
  \begin{equation}
  \label{eq_nu_eq1}
  \frac{dN}{dp_{T}} = a \times \frac{p_{T}}{(1+b^{2}p_{T}^{2})^{n}}
  \end{equation}
    \begin{equation}
  \label{eq_nu_eq2}
 \frac{dN}{dp_{T}} = l \times p_{T} \times \exp^{-\frac{m_{T}}{h}},\qquad m_{T} = \sqrt{p_{T}^{2}+m_{J/\psi}^{2}} 
  \end{equation}
  where $a$, $b$, $n$, $h$ and $l$ are free parameters. The difference between these two functional fits has been taken as a source of systematic uncertainty. Note that $R_{CP}$ reflects only relative suppression - if the modification of $J/\psi$ yield in central and peripheral bins is the same, $R_{CP}$ is equal to 1.   The $R_{CP}$, as a function of the average number of participant nucleons ($\langle N_{part}\rangle$), for Au+Au collisions at $\sqrt{s_{NN}}$ = 39, 62.4 and 200 GeV, are shown in Fig.~\ref{rcp_Fig}.  Note that the peripheral bin selection is 40 - 60\% central Au+Au collisions for these three energies.  The systematic uncertainties for $R_{CP}$ are mainly from TPC tracking cuts. Systematic uncertainties originating from  yield extraction, BEMC and TOF related cuts, and $n\sigma_{e}$ cuts, are negligible or mostly cancel. Significant suppression is observed in central Au+Au collisions at 62.4 GeV, which is similar to 200 GeV. 
  
  $R_{AA}$ is obtained {\color{black}from} comparing $J/\psi$ production in A+A collisions to  $p+p$  collisions, defined as follows:\\
  \begin{equation}
  \label{eq2}
  R_{AA}=\frac{1}{T_{AA}} \frac{d^{2}N_{AA}/dp_{T}dy}{d^{2}\sigma_{pp}/dp_{T}dy}
  \end{equation}
\\where $d^{2}N_{AA}/dp_{T}dy$ is the $J/\psi$ yield in A+A collisions and $d^{2}\sigma_{pp}/dp_{T}dy$ is the $J/\psi$ cross section in $p+p$  collisions. The nuclear overlap function is defined as $T_{AA} = \langle N_{coll}\rangle/\sigma^{pp}_{inel}$, where  $\sigma^{pp}_{inel}$ is the inelastic cross section in $p + p$ collisions and is equal to $34 \pm 3, 36 \pm 3$ and $42 \pm 3$ mb for 39, 62.4 and 200 GeV ~\cite{pp_cross}, respectively.  If there are no  hot or cold nuclear matter effects, the value of  $R_{AA}$ should be unity.

To obtain $R_{AA}$ at $\sqrt{s_{NN}}$ = 39 and 62.4 GeV, we have to derive the $J/\psi$ cross section in $p+p$ collisions because there are no measurements available for the $p+p$ references at STAR for these  two energies. There are several $p+p$  measurements from fixed target $p+$A experiments ~\cite{Alexopoulos_Phys,Schub_Phys,Gribushin_Phys} and from Intersecting Storage Ring (ISR) collider experiments ~\cite{Clark_Nucl,Kourkounelis_Phys} near these two energies. However, the $p_{T}$ shapes from Ref.~\cite{Clark_Nucl} and Ref.~\cite{Kourkounelis_Phys} at 63 GeV are inconsistent with each other and the cross section measurements at  39 GeV are comparable to (or even larger than) that at 63 GeV. Therefore, we use the cross section derived in  Ref.~\cite{Nelson_arxiv}  as our $p+p$ reference baselines for $\sqrt{s_{NN}}$= 39 and 62.4 GeV. In Ref.~\cite{Nelson_arxiv}, the world-wide experimental data on $J/\psi$ cross sections and kinematic distributions in $p+p$ and $p+$A collisions at $\sqrt{s}$ = 6.8 - 7000 GeV are examined in a systematic way. The authors explore the  $\sqrt{s}$ dependence of the inclusive cross section, rapidity and transverse momentum distributions phenomenologically and develop a strategy for the interpolation of the $J/\psi$ cross section and kinematics at RHIC energies. This approach is found to describe the world-wide $J/\psi$ data reasonably well. With this strategy, the predicted $J/\psi$ cross section times branching ratio at $\sqrt{s} =$ 39 and 62.4 GeV in mid-rapidity are ${\color{black}\text{Br}(J/\psi \rightarrow e^{+}e^{-})}d\sigma/dy|_{|y|<1.0} = 9.0 \pm 0.6$ and $17.6 \pm 2.1$ nb, respectively.

\renewcommand{\floatpagefraction}{0.75}
\begin{figure*}[htbp]
\includegraphics[angle=0,keepaspectratio,width=1.0\textwidth]{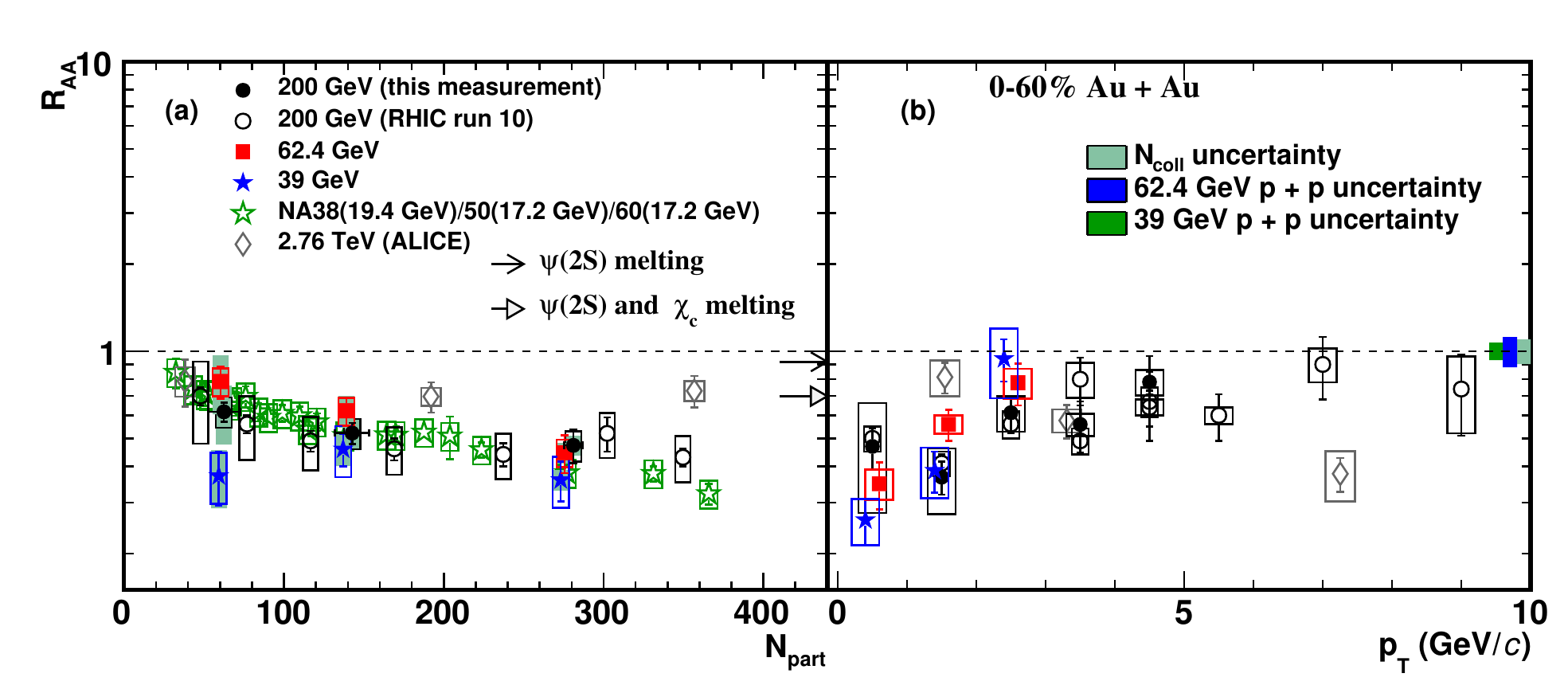}
\caption{The results of $J/\psi$ $R_{AA}$ as a function of $N_{part}$ (a) and $p_{T}$ (b) in Au+Au collisions at $\sqrt{s_{NN}}$  = 39, 62.4 and 200 GeV. The error bars represent the statistical uncertainties. The boxes represent the systematic uncertainties. The shaded bands indicate the uncertainties from $\langle N_{coll}\rangle$ and the uncertainties for the derived baselines for 39 and 62.4 GeV~\cite{Nelson_arxiv}.  The ALICE points are from ~\cite{alice_data}. The ratio of  feed-down  $J/\psi$ from higher chamonium states to inclusive $J/\psi$  is from ~\cite{feed-down}. The results of "RHIC run 10" are from ~\cite{jpsi_zebo} and ~\cite{jpsi_chris}.}
\label{raa_Fig}
\end{figure*}
\renewcommand{\floatpagefraction}{0.75}
\begin{figure*}[htbp]
\includegraphics[angle=0,keepaspectratio,width=1.0\textwidth]{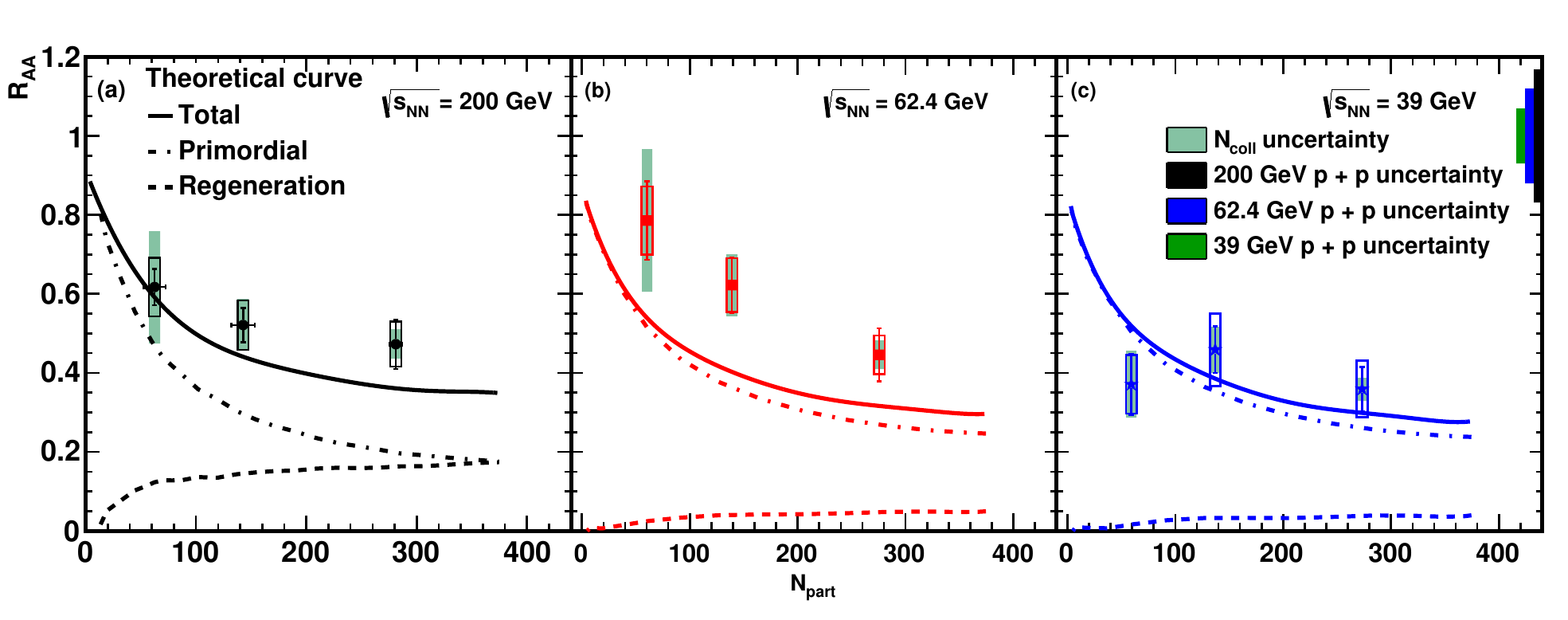}
\caption{The results of $J/\psi$ $R_{AA}$ as a function of $N_{part}$, in comparison with model calculations ~\cite{Zhao_Phys}, for Au+Au collisions at $\sqrt{s_{NN}}$  = 200 (a), 62.4 (b) and 39 GeV (c), respectively. The error bars represent the statistical uncertainties. The boxes represent the systematic uncertainties. The shaded bands indicate the uncertainties from $\langle N_{coll}\rangle$ and the uncertainties in the derived baselines for 39 and 62.4 GeV~\cite{Nelson_arxiv}. Solid lines are $J/\psi$ modification factors from model; dash-dotted line are suppressed primordial production; dashed line are regeneration component.}
\label{raa_forward_Fig}
\end{figure*}
With the derived $p+p$ references for 39 and 62.4 GeV, and the measured $p+p$ baseline at 200 GeV ~\cite{jpsi_zebo, reference_phenix}, we obtain the $R_{AA}$ of $J/\psi$ for $p_T\!>0$ as a function of $N_{part}$ in Au+Au collisions at $\sqrt{s_{NN}}$ = 39, 62.4, and 200 GeV, as shown in Fig.~\ref{raa_Fig} (a).
 The {\color{black}differential $R_{AA}$ in $J/\psi$ $p_T$} is shown in Fig.~\ref{raa_Fig} (b). The measurements from SPS ~\cite{Arnaldi_Phys,SPS_jpsi1,SPS_jpsi2} and LHC ~\cite{alice_data} and the expected $R_{AA}$ with complete $\psi(2S)$ and $\chi_{c}$ melting and no modification of the $J/\psi$ yield ~\cite{feed-down} are also {\color{black}included} for comparison. Suppression of $J/\psi$ production is observed in Au+Au collisions from 39 to 200 GeV with respect  to the production in $p+p$ scaled by $N_{coll}$. For $R_{AA}$ as a function of $N_{part}$, no significant energy dependence is observed within uncertainties from 17.2 to 200 GeV. For the $J/\psi$ $R_{AA}$ as a function of $p_{T}$, significant suppression is observed at low $p_{T}$ ($p_{T} < $ 2 GeV/$c$) from 39 to 200 GeV. The modification of $J/\psi$ production is consistent within the systematic uncertainties for these collision energies. The ALICE ~\cite{alice_data} points are also shown for comparison. In comparison with PHENIX results at forward rapidity ~\cite{Jpsi_phenix}, the suppression of $J/\psi$ shows no rapidity dependence at $\sqrt{s_{NN}} = $ 39 nor 62.4 GeV within uncertainties.  
 
\begin{figure*}[htbp]
\includegraphics[angle=0,keepaspectratio,width=0.6\textwidth]{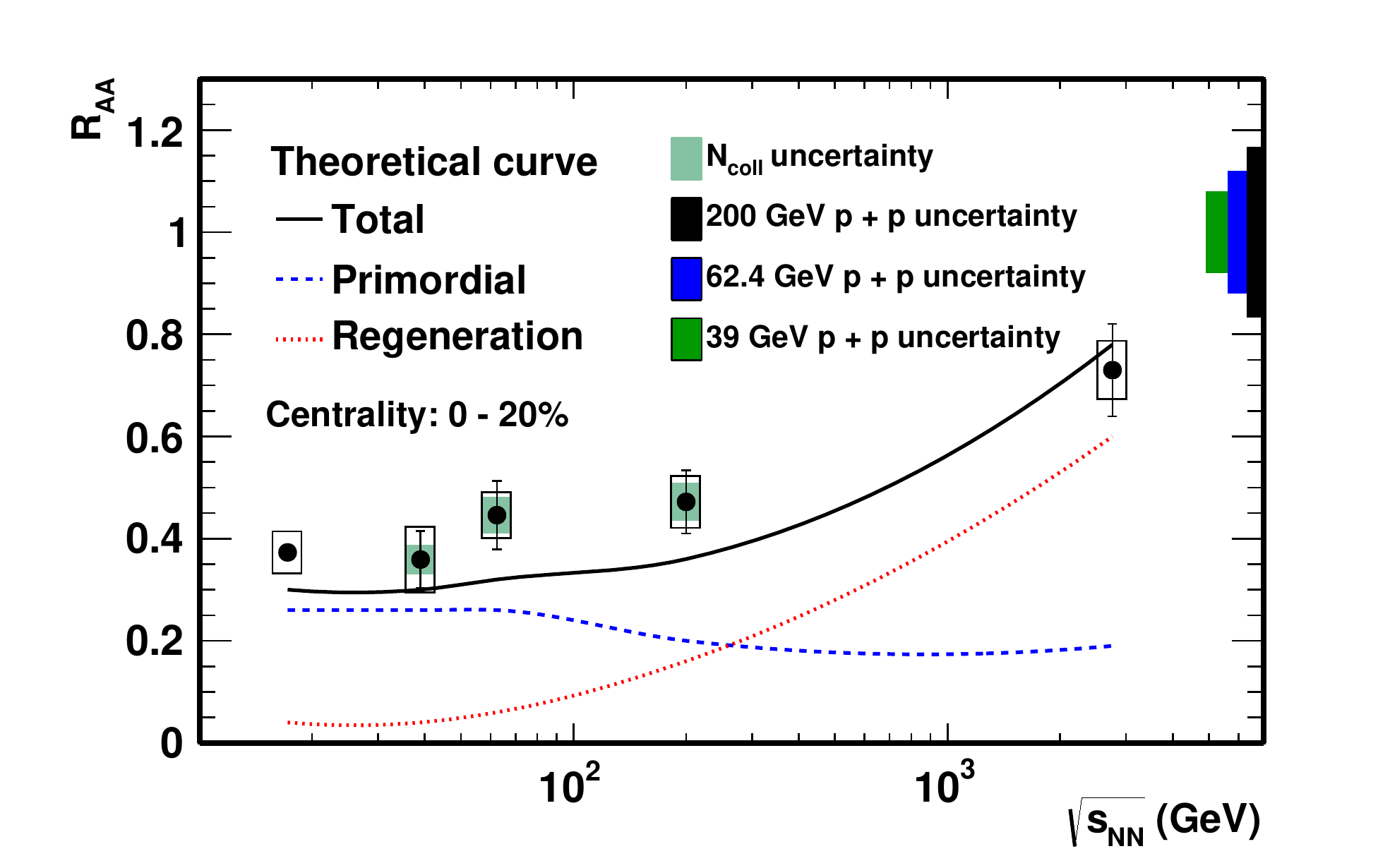}
\caption{The results of $J/\psi$ $R_{AA}$ as a function of collision energy for centrality 0-20 $\%$, in comparison with model calculations ~\cite{Zhao_Phys}. The SPS result ($\sqrt{s_{NN}}$ = 17.2 GeV) is from ~\cite{SPS_jpsi2}; the ALICE point ($\sqrt{s_{NN}}$ = 2.76 TeV) is from ~\cite{alice_data}. The error bars represent the statistical uncertainties and the boxes represent the systematic uncertainties. The shaded bands indicate the uncertainties from $\langle N_{coll}\rangle$ and the uncertainties for the derived baselines for 39 and 62.4 GeV~\cite{Nelson_arxiv}. Solid line is the total $J/\psi$ modification factors from model; dash-dotted line is the suppressed primordial production; dashed line is the regeneration component. Note: the ALICE point, $\sqrt{s_{NN}}$ =  2.76 TeV, in this figure is for 0-10 $\%$ centrality.}
\label{raa_energy_Fig}
\end{figure*}

 As shown in Fig.~\ref{raa_forward_Fig}, theoretical calculations ~\cite{Zhao_Phys} with initial suppression and $J/\psi$ regeneration describe the data within 1.6 standard deviation. The $R_{AA}$ results as a function of collision energy for 0-20 $\%$ centrality are also shown in Fig.~\ref{raa_energy_Fig}. Since ALICE data show no significant centrality dependence, we think it is appropriate to use the available 0-10$\%$ data at 2.76 TeV ~\cite{alice_data}. Theoretical calculations are also included for comparison. The calculations include two components: direct suppression and regeneration. The direct suppression represents the ''anomalous'' suppression of primordial $J/\psi$s due to CNM and color screening effects. According to the model calculations, the $R_{AA}$ is about 0.6 for central collisions with only CNM effects. The regeneration component is responsible for the contribution from the recombination of correlated or uncorrelated $c\bar{c}$ pairs.  The feed-down to $J/\psi$ from $\chi_{c}$ and $\psi^{\prime}$ has been taken into account in the calculations. No significant energy dependence of $R_{AA}$ for 0-20 $\%$ centrality is observed at $\sqrt{s_{NN}} <$ 200 GeV.  As the collision energy increases the QGP temperature increases, thus the $J/\psi$ color screening becomes more significant. However, in the theoretical calculation~\cite{Zhao_Phys}, the regeneration contribution increases with collision energy due to the increase in the charm pair production, and nearly compensates the enhanced suppression arising from the higher temperature. The higher $R_{AA}$ at ALICE may indicate that the surviving $J/\psi$s are mainly coming from the recombination contribution. The model calculation describes the energy dependence of $J/\psi$ production from SPS to LHC.

\renewcommand{\floatpagefraction}{0.75}

\section{Summary}
In summary, we report on recent STAR measurements of  $J/\psi$ production at midrapidity in Au+Au collisions at $\sqrt{s_{NN}}$ = 39, 62.4 and 200 GeV. Suppression of $J/\psi$ production, with respect to the production in $p+p$ scaled by the number of binary nucleon-nucleon collisions, is observed at these three energies. The observed suppression is consistent with the suppression of directly produced $J/\psi$ mesons. No significant energy dependence of the nuclear modification factor (either $R_{AA}$ or $R_{CP}$) is found within uncertainties. Model calculations, which include direct suppression and regeneration, reasonably describe the centrality and energy dependence of $J/\psi$ production in high-energy heavy ion collisions.

\section{Acknowledgments}
We thank the RHIC Operations Group and RCF at BNL, the NERSC Center at LBNL, the KISTI Center in
Korea, and the Open Science Grid consortium for providing resources and support. This work was 
supported in part by the Office of Nuclear Physics within the U.S. DOE Office of Science,
the U.S. NSF, the Ministry of Education and Science of the Russian Federation, NSFC, CAS,
MoST and MoE of China, the National Research Foundation of Korea, NCKU (Taiwan), 
GA and MSMT of the Czech Republic, FIAS of Germany, DAE, DST, and UGC of India, the National
Science Centre of Poland, National Research Foundation, the Ministry of Science, Education and 
Sports of the Republic of Croatia, and RosAtom of Russia.


\begin{thebibliography}{9}

\bibitem{Harrison_Nucl} M. Harrison et al., Nucl. Instr. Meth. A \textbf{499}, 235 (2003).
\bibitem{Matsui_Phys} T. Matsui and H. Satz, Phys. Lett. B \textbf{178}, 416 (1986).
\bibitem{Lansberg_Mod} J. P. Lansberg, Int J. Mod. Phys. A \textbf{21}, 3857 (2006).
\bibitem{Abreu_phys} M. C. Abreu et al., Phys. Lett. B \textbf{449}, 128 (1999).
\bibitem{Arnaldi_Phys} R. Arnaldi et al., Phys. Rev. Lett \textbf{99}, 132302 (2007).
\bibitem{Atomssa_Eur} E. T. Atomssa, Eur. Phys. J. C \textbf{61}, 683 (2009).
\bibitem{SPS_jpsi1} M.C. Abreu et al., Phys. Lett. B \textbf{466}, 408 (1999).
\bibitem{SPS_jpsi2} M.C. Abreu et al., Phys. Lett. B \textbf{477}, 28 (2000).
\bibitem{Adare_Phys} A. Adare et al., Phys. Rev. Lett. \textbf{98}, 232301 (2007).
\bibitem{jpsi_chris} L. Adamczyk et al., Phys. Rev. C \textbf{90}, 024906 (2014).
\bibitem{energy_density} K. J. Eskola et al., Nucl. Phys. B \textbf{570}, 379 (2000).
\bibitem{10_zebo} L. Grandchamp, R. Rapp and G.E. Brown, Phys. Rev. Lett. \textbf{92}, 212301 (2004).
\bibitem{Zhao_Phys} X. Zhao and R. Rapp, Phys. Rev. C \textbf{82}, 064905 (2010) (private communication).

\bibitem{Hot_2} R. Rapp et al., Prog. Part. Nucl. Phys. \textbf{65}, 209 (2010).
\bibitem{Hot_3} Y. Liu et al., Int. J. Mod. Phys. E \textbf{24}, 1530015 (2015).

\bibitem{CNM_1} J. L. Nagle et al., Phys. Rev. C \textbf{84}, 044911 (2011).
\bibitem{CNM_2} J. W. Cronin et al., Phys. Rev. D \textbf{11}, 3105 (1975).
\bibitem{CNM_3} J. L. Nagle and M. J. Bennett, Phys. Lett. B \textbf{465}, 21 (1999).
\bibitem{CNM_4} R. Vogt, Nucl. Phys. A \textbf{700}, 539 (2002).
\bibitem{CNM_5} E. G. Ferreiro, Phys. Lett. B \textbf{749}, 98 (2015).

 \bibitem{CNME_1} D. Alde et al., Phys. Rev. Lett. \textbf{66}, 133 (1991).
 \bibitem{CNME_2} M. Leitch et al., Nucl. Phys. A \textbf{544}, 197c (1992).
 \bibitem{CNME_3} M. Leitch  et al., Phys. Rev. Lett. \textbf{84}, 3256 (2000).
 \bibitem{CNME_4} B. Alessandro et al., Eur. Phys. J. C \textbf{33}, 31 (2004).
  \bibitem{CNME_5} B. Alessandro et al., Eur. Phys. J. C \textbf{39}, 335 (2005).
 \bibitem{CNME_6} R. Arnaldi et al., Phys. Lett. B \textbf{706}, 263 (2012).
 \bibitem{CNME_7} A. Adare et al., Phys. Rev. C \textbf{87}, 034904 (2013).
  \bibitem{CNME_8} A. Adare et al., Phys. Rev. Lett. \textbf{107}, 142301 (2011)
 \bibitem{CNME_9} A. Adare et al., Phys. Rev. Lett. \textbf{111}, 202301 (2013).   

\bibitem{Jpsi_ALICE} B. Abelev et al., Phys. Rev. Lett. \textbf{109}, 072301 (2012).
\bibitem{Jpsi_ATLAS} G. Aad et al., Phys. Lett. B \textbf{697}, 294 (2011).
\bibitem{Jpsi_CMS} S. Chatrchyan et al., JHEP \textbf{05}, 063 (2012).
\bibitem{Jpsi_phenix} A. Adare et al., Phys. Rev. C \textbf{86}, 064901 (2012).
\bibitem{Ackermannn_Nucl} K. H. Ackermannn et al., Nucl. Instr. Meth. A \textbf{499}, 624 (2003).
\bibitem{VPD} W. J. Llope et al., Nucl. Instr. Meth. A \textbf{759}, 23 (2014).
\bibitem{PDG} K. A. Olive et al., Chin. Phys. C\textbf{38}, 090001 (2014).
\bibitem{Anderson_Nucl} M. Anderson et al., Nucl. Instr. Meth. A  \textbf{499}, 659 (2003).
\bibitem{Llope_Nucl} W. J. Llope, Nucl. Instr. Meth. A  \textbf{661}, S110 (2012).
\bibitem{Beddo_Nucl} M. Beddo et al., Nucl. Instr. Meth. A  \textbf{499}, 725 (2003).
\bibitem{jpsi_zebo}L. Adamczyk et al., Phys. Lett. B \textbf{722}, 55 (2013).
\bibitem{Glauber_model} M.L. Miller et al., Ann. Rev. Nucl. Part. Sci. \textbf{57}, 205 (2007).
\bibitem{19_zebo}  B. I. Abelev et al., Phys. Rev. C  \textbf{79}, 034909 (2009).
\bibitem{dielectron} L. Adamczyk et al., Phys. Rev. C \textbf{92}, 024912 (2015).
\bibitem{pp_cross} A. Achilli et al., Phys. Rev. D \textbf{84}, 094009 (2011).


\bibitem{Alexopoulos_Phys} T. Alexopoulos et al., Phys. Rev. D  \textbf{55}, 3927 (1997).
\bibitem{Schub_Phys} M. H.  Schub et al., Phys. Rev. D \textbf{52}, 1307 (1995).
\bibitem{Gribushin_Phys} A. Gribushin et al., Phys. Rev. D  \textbf{62}, 012001 (2000).
\bibitem{Clark_Nucl} A. G. Clark et al., Nucl. Phys. B  \textbf{142}, 29 (1978).

\bibitem{Kourkounelis_Phys} C. Kourkoumelis et al., Phys. Lett. B  \textbf{91}, 481 (1980).
\bibitem{Nelson_arxiv} W. Zha et al., Phys. Rev. C \textbf{93}, 024919 (2016).
\bibitem{reference_phenix} A. Adare et al., Phys. Rev. D \textbf{82}, 012001 (2010).
\bibitem{alice_data} B. Abelev et al., Phys. Lett. B \textbf{734}, 314 (2014).
\bibitem{feed-down} P. Faccioli et al., JHEP \textbf{10}, 004 (2008).

\expandafter\ifx\csname
natexlab\endcsname\relax\def\natexlab#1{#1}\fi
\expandafter\ifx\csname bibnamefont\endcsname\relax
  \def\bibnamefont#1{#1}\fi
\expandafter\ifx\csname bibfnamefont\endcsname\relax
  \def\bibfnamefont#1{#1}\fi
\expandafter\ifx\csname citenamefont\endcsname\relax
  \def\citenamefont#1{#1}\fi
\expandafter\ifx\csname url\endcsname\relax
  \def\url#1{\texttt{#1}}\fi
\expandafter\ifx\csname
urlprefix\endcsname\relax\def\urlprefix{URL }\fi
\providecommand{\bibinfo}[2]{#2}
\providecommand{\eprint}[2][]{\url{#2}}


\end{thebibliography}
\end{document}